\documentclass[12pt,preprint]{aastex}

\def\ltsima{$\; \buildrel < \over \sim \;$}
\def\gtsima{$\; \buildrel > \over \sim \;$}
\def\lsim{\lower.5ex\hbox{\ltsima}}
\def\gsim{\lower.5ex\hbox{\gtsima}}
\def\lapp{\ifmmode\stackrel{<}{_{\sim}}\else$\stackrel{<}{_{\sim}}$\fi}
\def\gapp{\ifmmode\stackrel{>}{_{\sim}}\else$\stackrel{<}{_{\sim}}$\fi}

\usepackage{amsmath}
\usepackage{textcomp}
\usepackage{amssymb}
\usepackage{multirow}
\usepackage{booktabs}
\usepackage{graphicx}

\newdimen\minuswidth    
\setbox0=\hbox{$-$}
\minuswidth=\wd0
\catcode`@=\active
\def@{\kern\minuswidth}
\setbox0=\hbox{\rm0}
 
\shorttitle{Optical counterpart to EXO 1745-248} 
\shortauthors{Ferraro et al.}
 
\begin{document} 
\title{Probing the MSP prenatal stage: the optical identification of
  the X-ray burster EXO 1745-248 in Terzan 5}

\author{
F. R. Ferraro\altaffilmark{1},
C. Pallanca\altaffilmark{1},
B. Lanzoni\altaffilmark{1},
M. Cadelano\altaffilmark{1,2},
D. Massari\altaffilmark{2,3},
E. Dalessandro\altaffilmark{1},
A. Mucciarelli\altaffilmark{1},
}

\affil{\altaffilmark{1} Dipartimento di Fisica e Astronomia,
  Universit\`a degli Studi di Bologna, v.le Berti Pichat 6/2,
  I$-$40127 Bologna, Italy}
\affil{\altaffilmark{2}INAF-Osservatorio Astronomico di Bologna, via
  Ranzani 1, 40127, Bologna, Italy}
\affil{\altaffilmark{3}Kapteyn Astronomical Institute, University of
  Groningen, PO Box 800, 9700 AV Groningen, The Netherlands}
\altaffiltext{$^\ast$}{Based on observations (GO 14061, GO 12933, GO 9799) with
  the NASA/ESA \textit{Hubble Space Telescope}, obtained at the Space
  Telescope Science Institute, which is operated by AURA, Inc., under
  NASA contract NAS 5-26555.}

\date{02 May, 2015}

\begin{abstract}
We report on the optical identification of the neutron star burster
EXO 1745-248 in Terzan 5.  The identification was performed by
exploiting HST/ACS images acquired in  Director's Discretionary
Time shortly after (approximately 1 month) the Swift detection of the
X-ray burst.  The comparison between these images and previous 
  archival data revealed the presence of a star that currently
brightened by $\sim 3$ magnitudes,  consistent with expectations
  during an X-ray outburst.   The centroid of this object well
  agrees with the position, in the archival images, of a star located
  in the Turn-Off/Sub Giant Branch region of Terzan 5.  This supports
  the scenario that the companion should has recently filled its Roche
  Lobe. Such a system represents the pre-natal stage of a
millisecond pulsar, an evolutionary phase during which heavy mass
accretion on the compact object occurs, thus producing X-ray outbursts
and re-accelerating the neutron star.
\end{abstract}
 

\section{INTRODUCTION}
Low mass X-ray binaries (LMXBs) and radio millisecond pulsars (MSPs)
are thought to be, respectively, the starting and the ending stages of
a common evolutionary path, where a neutron star accretes matter (and
angular momentum) from a companion (e.g., \citealp{bhattacharya91,
  wijnands98}).  The early phases of this evolutionary path are
characterized by active mass accretion accompanied by intense X-ray
emission (larger than $\sim 10^{35}$ erg s$^{-1}$). These systems are
observed as LMXBs characterized by a few outburst in the X-ray due to
accretion disk instabilities.  These objects are usually called
``X-ray transients'' \citep{white84}.  When the neutron star is
re-accelerated, the system will appear as a MSP in the radio band.
Newly born MSPs are therefore expected to have a bloated and tidally
deformed companion which is still losing mass from its Roche Lobe. In
this case, the system is called ``redback'' (or ``black widow'' if the
companion is less massive than $0.05 M_\odot$).  Following the
canonical scenario, at the end of the evolution only the degenerate
core of the peeled companion (i.e., a helium or carbon-oxygen white
dwarf) is predicted to orbit the MSP (a sub-stellar mass object, which
is eventually completely evaporated, in the case of a black widow).
Indeed, the optical searches for the companion stars to binary MSPs
performed so far in Galactic globular clusters detected objects
belonging to all these three classes: five bloated stars, companions
to redbacks (see \citealp{fe01, edmonds02, cocozza08, pallanca10,
  pallanca13M28}), two very low mass objects companions to black
widows (\citealp{pallanca14}, Cadelano et al. 2015, submitted to ApJ),
and eight white dwarfs companions to canonical MSPs
(\citealp{edmonds01,bassa03,fe036752,sigurdsson03}, Mario Cadelano et
al. in preparation, Liliana Rivera Sandoval et al. in preparation).  A
new unexpected link has been added to the chain very recently.
\citet{papitto13} found that the X-ray transient IGR J18245-2452 in
the globular cluster M28 \citep{eckert13} corresponds to MSP M28I, and
the system is currently swinging between accretion-powered and
rotation-powered states. This shows that the transition from a LMXB to
a MSP passes through an intermediate phase during which the two states
cyclically alternate over a time scale of a few years
\citep{papitto13}.  A low-mass main sequence star, which experienced a
strong ($\sim 2$ magnitudes) luminosity increase, has been identified
as the optical counterpart to the system \citep{pallanca13M28}.

On March 13, 2015, Swift/BAT observations detected an X-ray burst in
Terzan 5 \citep{altamirano15}.  The Swift/XRT observations promptly
following the Swift/BAT detection localized the transient source at
RA(J2000)$=267.0207\deg$, DEC(J2000)$=-24.779\deg$, with a 90\%
uncertainty of $3.5\arcsec$ \citep{bahramian15}.  Moreover, the
measured spectrum turned out to be consistent with a relatively hard
photon index of $1.0 \pm 0.2$ and a hydrogen column density $N_{\rm H}
=(4\pm 0.8)\times 10^{22}$ cm$^{-2}$. The latter is larger than the
typical value measured in Terzan 5 \citep{bahramian14} and well in
agreement with the hydrogen column density of the previously known
transient EXO 1745-248 \citep{kuulkers03}.  Indeed, the subsequent
position refinement by \citet{linares15} centered the system around
EXO 1745-248 with a $2.2\arcsec$ error circle. These data therefore
strongly suggest that the new Swift/BAT outburst coincides with EXO
1745-248, an X-ray neutron star transient that already showed
outbursts in 2000 and 2011 \citep{degenaar12}.  Such an identification
has been also confirmed by radio VLA observations \citep{tremou15},
which locate the source position within $0.4\arcsec$ of the published
coordinates of EXO 1745-248 obtained from \emph{Chandra} data (source
CX3 in \citealp{heinke06}).  The most recent Swift/XRT observations
indicate that the source is probably on the way to transit to the soft
state \citep{yan15}.

Such an intriguing object is not uncommon in Terzan 5. In fact, this
stellar system is known to harbor several X-ray sources (see, e.g.,
\citealp{heinke06}) and to be the most efficient furnace of MSPs in
the Milky Way: it harbors a total of 34 MSPs, corresponding to $\sim
25\%$ of the entire sample of such objects known to date in Galactic
globular clusters \citep{ransom05}.  \citet{fe09} recently
demonstrated that, at odds with what is commonly thought, Terzan 5 is
not a globular cluster, but a system hosting stellar populations
characterized by significantly different iron abundances, spanning a
total metallicity range of 1 dex (see also \citealp{ori11, ori13,
  massari14}).  The measured chemical patterns \citep{ori11, ori13}
could be naturally explained in a scenario where Terzan 5 was
originally much more massive than today ($\sim 10^6 M_\odot$;
\citealp{lan10}), thus to be able to retain the iron-enriched gas
ejected by violent supernova explosions.  The large number of type II
supernovae required to explain the observed abundance patterns should
have also produced a large population of neutron stars, mostly
retained into the deep potential well of the massive {\it
  proto}-Terzan 5 and likely forming binary systems through tidal
capture interactions.  Finally, its large collision rate
\citep{lan10}, the largest among all Galactic globular clusters
\citep[see also][]{verhut87}, could have highly promoted pulsar
re-cycling processes, which can explain the production of the large
population of MSPs/LMXBs now observed in the system.

Since  the X-ray outburst detected by SWIFT is expected to also
produce a significant enhancement of the optical luminosity
  \citep[see][]{shahbaz98,charles06,testa12, pallanca14}, we
successfully applied for HST Director Discretionary Time to urgently
survey the central region of Terzan 5 and thus provide new insights
into this still unexplored phase of the LMXB-to-MSP path. Here we
report on the identification of the optical counterpart to EXO
1745-248 obtained from the analysis of these images.

\section{OBSERVATIONS AND DATA REDUCTION}
\label{obs}
To search for the expected optical emission from EXO 1745-248 during
its X-ray bursting phase, we submitted a HST Director Discretionary
Time proposal (GO 14061, PI: Ferraro) asking for two orbits with the
Advanced Camera for survey (ACS/WFC).  The observations have been
promptly performed on April 20, 2015, about one month  into the
  X-ray outburst (continuing at time of writing).  The dataset
(hereafter Epoch 3, EP3) consists of $5\times398$ s images in F606W,
$5\times371$ s images in F814W, and one short exposure per filter
($50$ s and $10$ s, respectively; these latter have not been used in
the present work).

Previous optical images of Terzan 5 acquired with the same instrument
in the same filters were already present in the HST Archive: GO 12933
(PI: Ferraro) performed on August 18th, 2013 (hereafter EP2), and
GO 9799 (PI: Rich) performed on September 9th, 2003 (hereafter EP1).
We already used these data to construct the deepest optical
color-magnitude diagram (CMD) of Terzan 5 (see \citealp{fe09, lan10,
  massari12}).

For the present study, all the datasets have been homogeneously
analyzed, applying standard IRAF procedures for Pixel-Area-Map (PAM)
correction on the (flc) images corrected for Charge Transfer
Efficiency.  The photometric analysis has been carried out by using
the DAOPHOT package.  For each image we modeled a Moffat point spread
function (PSF) by using 150-200 bright and nearly isolated stars.
Afterwards, we performed source detection in each image imposing a
3-$\sigma$ threshold over the background level.  By using all the
sources detected and PSF fitted in at least 1 out of 2 images in EP1,
and 3 out of 10 images in EP2 and EP3, we then created a catalog for
every epoch.  In spite of including sources detected in only one
filter, such an approach allowed us to avoid losing very faint (but
possibly real) objects, while safely discarding spurious detections,
as cosmic rays and detector artifacts.  The obtained master lists (one
for every epoch) have been then used to identify the stellar sources
in each single frame and the PSF model has been applied to derive the
final magnitudes. As a final step, to build the EP2 and EP3 catalogs
we considered all the stars with magnitude measured in both filters in
at least 3 out of 5 images, while the EP1 catalog obviously consists
of the objects detected in both the available images.  For each star,
the magnitudes estimated in different images of the same filter have
been homogenized \citep[see][]{fe92} and their weighted mean and
standard deviation have been finally adopted as the star magnitude and
photometric error.  The magnitude calibration to the VEGAMAG system
has been performed by using the catalog by \citet{massari12} as
reference.

To precisely determine the star coordinates, we first applied the
equations reported by \citet{meurer02} and corrected the instrumental
positions for the known geometric distortions affecting the ACS
images. Through cross-correlation with the catalog of
\citet{massari12}, which had been placed onto the 2MASS system, we
then obtained the absolute coordinates for each star, with a final
astrometric accuracy of $\sim 0.2\arcsec$ in both right ascension and
declination.

\begin{figure}[!htb]
\centering%
\includegraphics[scale=0.9]{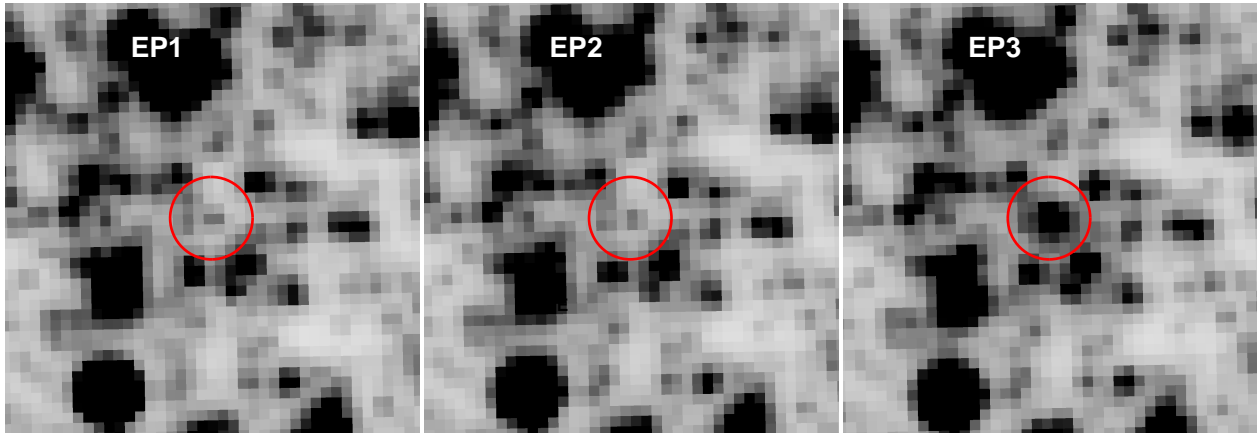}
\caption{HST/ACS drz combined images of the $2\arcsec\times 2\arcsec$
  region around EXO 1745-248, in the F814W filter, for the three
  epochs (EP1, EP2, EP3, from left to right, respectively). The source
  (highlighted with a red circle) is visible as a faint star during
  the quiescence epochs EP1 and EP2, while it is observed in an
  outburst stage during EP3. North is up, east is to the left.}
\label{ident1}
\end{figure}

\begin{figure}[!htb]
\centering%
\includegraphics[scale=0.7]{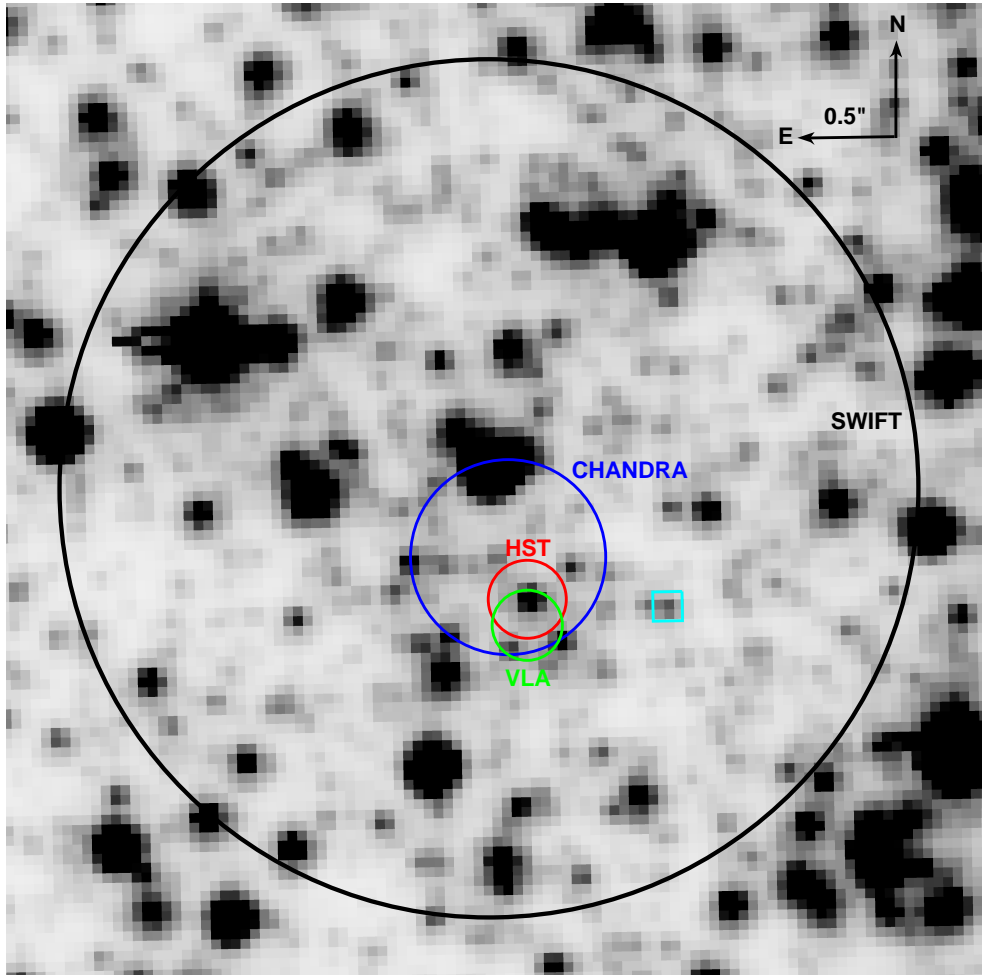}
\caption{F814W-band drz combined image of the $5\arcsec\times
  5\arcsec$ region around EXO 1745-248 in the EP3 exposure. The source
  positions and uncertainties obtained from the various observational
  campaigns are marked: the Swift/XRT $2.2\arcsec$ radius error circle
  is shown in black, the Chandra error circle in blue, the VLA measure
  in green, and the HST optical determination in red. The cyan square
  marks the star previously proposed \citep{heinke03} as the possible
  optical counterpart to EXO 1745-248.}
\label{ident2}
\end{figure}

\begin{figure}[!htb]
\centering%
\includegraphics[scale=0.7]{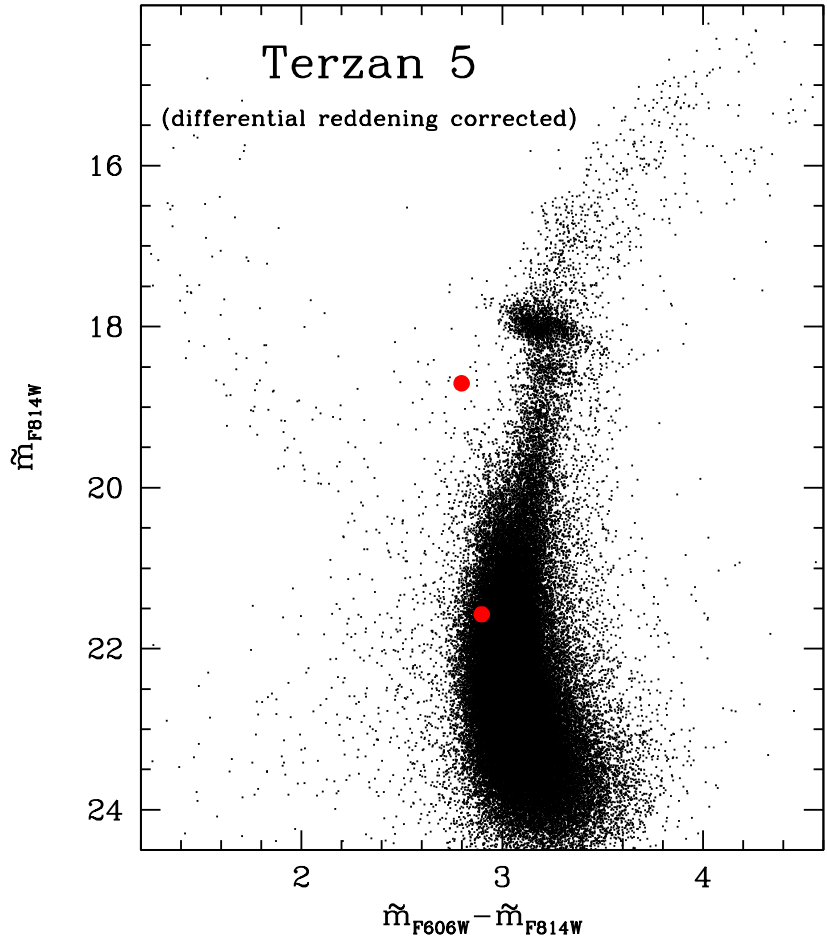}
\caption{(m$_{\rm F814W}$, m$_{\rm F606W}-$m$_{\rm F814W}$) CMD of
  Terzan 5 corrected for differential reddening (according to
  \citealp{massari12}).  The position of the optical counterpart to
  EXO 1745-248, in the outburst and in quiescence states, is marked
  with large red circles.}
\label{cmd1}
\end{figure}

\begin{figure}[!htb]
\centering%
\includegraphics[scale=0.7]{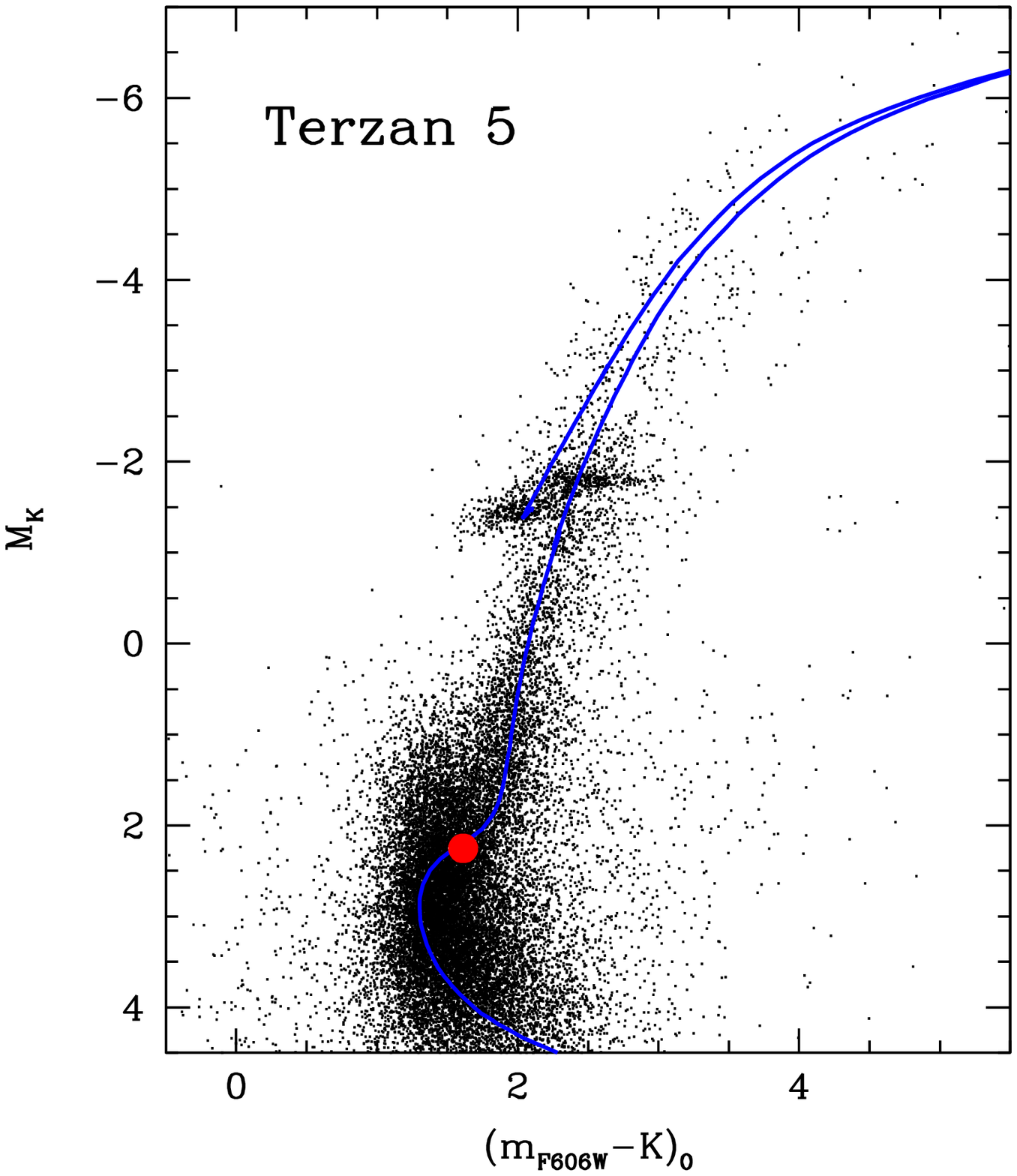}
\caption{Absolute $(M_K,$ m$_{\rm F606W}- K)_0$ CMD of Terzan 5
  obtained from a combination of HST/ACS and ESO/MAD observations. The
  position of COM-EXO 1745-248 in the quiescent state is marked with
  the large red circle. The blue line corresponds to a 12 Gyr
  isochrone with [Fe/H]$=-0.3$ (from \citealp{girardi10}), well
  reproducing the main metal poor sub-population of Terzan 5.
}
\label{cmd2}
\end{figure}
 
\section{RESULTS}
\label{results}
The photometric analysis of our dataset in a region around the
position of EXO 1745-248 immediately revealed, in EP3, the presence of
 a bright star that was not visible in EP1 and EP2 images (see
  Figure \ref{ident1}).  The comparison of the three epochs
  unequivocally identifies the bright object (hereafter COM-EXO
1745-248) as the optical counterpart to EXO 1745-248.  The absolute
position of the optical source is RA(J2000)$=17^h 48^m 05.23^s$,
Dec(J2000)$=-24^o 46\arcmin 47.6\arcsec$.  This is consistent at
1-$\sigma$ with the VLA position quoted by \citet[][see the red and
  the green circles in Figure \ref{ident2}]{tremou15}.  Instead, the
star previously suggested as the possible optical counterpart to this
X-ray transient \citep{heinke03} is located $\sim 0.7\arcsec$ to the
west (cyan square in the figure).

 After the astrometric transformations, the centroid position of
  the bright object in EP3 is within 0.05 pixels from the centroid of
  a fainter star clearly detected in EP1 and EP2 (see Figure
  \ref{cmd1}). This could be the optical counterpart caught in
  quiescence. The probability that the true counterpart is a fainter,
  non detected star aligned (within 0.05 pixels) along the line of
  sight is very low ($P \sim 0.4\%$).\footnote{To estimate the
    probability of a chance superposition with a star fainter than the
    proposed counterpart, the number of stars down to 5 magnitudes
    below the Turn Off level at the same distance ($\sim 5\arcsec$)
    from the cluster center is needed. Since no data-set available for
    Terzan 5 reaches such a faint magnitude limit, we adopted as
    reference the luminosity function of 47 Tucanae derived from deep
    HST observations \citep{sarajedini07}. Star counts have been
    normalized to the number of Terzan 5 stars counted between the
    Turn Off level and two magnitudes above, in a ring of $2\arcsec$
    width, centered at $5\arcsec$ from the center.} In addition, the
  brightness profile of this star does not show any significant
  deviation from symmetry, thus supporting the hypothesis that it is a
  single object. Hence, the most natural conclusion is that identified
  star is indeed the counterpart in quiescence.  The identified
object passed from observed magnitudes m$_{\rm F606W}=24.74$ and
m$_{\rm F814W}=21.74$ during quiescence (EP1 and
EP2)\footnote{The magnitudes of the star in the EP1 ($m_{\rm
      F606W}=24.7\pm0.1$; $m_{\rm F814W}=21.6\pm0.1$) and EP2 ($m_{\rm
      F606W}=24.74\pm0.04$; $m_{\rm F814W}=21.7\pm0.1$) quiescent
    stages are fully consistent within the errors.}, to m$_{\rm
  F606W}=21.77$ and m$_{\rm F814W}=18.88$ in the outburst state (EP3),
thus experiencing a brightening of 3 magnitudes (corresponding to a
factor 16 in luminosity).  Because of its location in the inner
Galactic bulge, Terzan 5 is affected by a large extinction, with an
average color excess $E(B-V)=2.38$ \citep{barbuy98, valenti07},
showing strong variations, up to $\delta E(B-V)=0.67$ mag, within the
ACS field of view \citep{massari12}. We therefore applied the
high-resolution differential reddening map obtained by
\citet{massari12} to correct the observed magnitudes  (in the
  following, the notation $\widetilde {\rm m}$ indicates magnitudes
  corrected for differential reddening). Figure \ref{cmd1} shows the
position of COM-EXO 1745-248 in the differential reddening corrected
CMD during the two states.  We found $\widetilde {\rm m}_{\rm
  F606W}=24.47$ and $\widetilde {\rm m}_{\rm F814W}=21.57$ during EP1
and EP2, while $\widetilde {\rm m}_{\rm F606W}=21.50$ and $\widetilde
{\rm m}_{\rm F814W}=18.70$ in EP3,  corresponding to a small (0.1
  mag) color variation, which is however within the errors.  No
variability has been detected over the period of $\sim 50$ min covered
by each HST orbit in EP3 and EP2. It is worth mentioning that EP3 data
were acquired on April 20, 2015, almost simultaneously to the X-ray
observations \citep{yan15} suggesting that the system is transiting
from a hard to a soft state.

An estimate of the orbital period of the system can be obtained by
following \citep{shahbaz98}, who reporte a relation between the
orbital period and the $V$-band luminosity variation. Since we observe
$\Delta V \sim 3$ mag in the case of EXO 1745-248, the orbital period
turns out to be $P\sim 1.3$ days. On the other hand, for LMXBs
\citet{vanpa94} proposed an empirical relation between the absolute
$V$ magnitude in outburst and the parameter $\Sigma$, which depends on
the ratio between the X-ray and the Eddington luminosities ($L_{\rm
  X}/L_{\rm Edd}$) and the orbital period.  By assuming $L_{\rm
  X}/L_{\rm Edd}\sim 0.5$ \citep{yan15} and $M_V =1.37$ (in Johnson
$V$ magnitude) for EXO 1745-248, we obtain $P\sim0.1$ days. From these
estimates, the orbital period of the system is likely to be between
1.3 and 0.1 days.

In order to more deeply investigate the nature of COM-EXO 1745-248 in
the quiescence state under the assumption that the disk contribution
to the observed magnitude is negligible, we identified the star in the
K-band adaptive optics images obtained with ESO/MAD, used by
\citet{fe09} to discover the two main multi-iron populations hidden in
this system.  We have first corrected the combined $(K,$ m$_{\rm
  F606W}-K$) CMD for differential reddening. Then we transformed it
into the absolute plane by assuming the average color excess quoted
above, and the distance modulus $(m-M)_0=13.87$ corresponding to a
distance of 5.9 kpc \citep{valenti07}.  The result is shown in Figure
\ref{cmd2}, where the position of COM-EXO 1745-248 in the quiescent
state is marked.  A more detailed characterization of the nature of
COM-EXO 1745-248 is strongly hampered by the complexity of the stellar
populations harbored in Terzan5. The comparison with a 12 Gyr old
isochrone \citep{girardi10} well reproducing the main metal-poor
sub-population of Terzan 5, at [Fe/H]$=-0.3$ dex\footnote{As discussed
  in \citet{massari14}, this population consists of $\sim 62\%$ of the
  total, while a super-solar component at [Fe/H]$=+0.3$ dex accounts
  for $\sim 29\%$, and an even metal poorer component, at
  [Fe/H]$=-0.8$ dex, recently detected by \citet{ori13}, corresponds
  to $\sim 5\%$ of the total. The CMD plotted in Fig. \ref{cmd2}
  nicely shows two distinct red clumps at $M_K=-1.5$ and $M_K=-1.8$1,
  corresponding to the two major sub-populations first discovered in
  the system by \citet{fe09}.} suggests that COM-EXO 1745-248 could be
a sub-giant branch (SGB) star. On the other hand the metal-rich
sub-population could be significantly (a few Gyr) younger than the
main metal poor component \citep[see][]{fe09}. Thus, if COM-EXO
1745-248 belongs to the metal-rich component, it would be located
below the SGB, in a position where companions to redback MSPs have
been found (see, e.g., the case of COM-MSP6397A in
\citealp{fe01}). Since no spectroscopic information on the metallicity
of this star is available, both possibilities are equivalently
valid. While in the case of redbacks any prediction on the stellar
parameters based on the observed photometric properties can be
difficult (see the case of COM-MSP6397A), this is possible for a SGB
star belonging to the metal poor population. In this case, the
following stellar parameters are obtained: mass $M=0.9 M_\odot$,
effective temperature $T_{\rm eff}= 5440$ K, surface gravity $\log
g=3.9$, and luminosity $\log L/L_\odot=0.35$. The corresponding
stellar radius therefore is $R\sim1.7 R_\odot$. Hence, by assuming
that the star has completely filled its Roche Lobe and adopting a
canonical value for the neutron star mass ($\sim 1.4 M_\odot$), we
derive an orbital separation $a \approx 5.2 R_\odot$ and a period
$P_{\rm orb}\sim 0.9$ d for the binary system, fully in agreement with
the range estimated above.  We estimate that the radial velocity
variations of such a binary system should have an amplitude of $\sim
170 \sin i$ km s$^{-1}$ ($i$ being the system inclination angle),
which could be detectable through a dedicated spectroscopic follow-up.

\section{DISCUSSION AND CONCLUSIONS}
\label{conclusions}

By using high-resolution images obtained with the HST/ACS during three
different epochs, we have identified the optical counterpart to the
neutron star transient EXO 1745-248 in Terzan 5.  With respect to the
two previous epochs, this object shows a current brightening of $\sim
3$ magnitudes.  In the quiescence state it is a sub giant branch star,
i.e., an object that is experiencing its first envelope expansion,
while evolving toward the red giant branch stage.

Very interestingly, the X-ray emission of EXO 1745-248 during
quiescence was found to be highly variable both on short and long
time-scales \citep{degenaar12}, and \citet{linares14} recently
underlined that these properties are impressively similar to those
observed for MSP-M28I, the system occasionally swinging between
accretion-powered and rotation-powered emission \citep{papitto13}.
This evidence suggests that EXO 1745-248 could be another system
belonging the rare class of objects caught shortly before the
formation of a radio MSP, possibly in a stage immediately preceding
the swinging phase in which MSP-M28I has been observed.

The characterization of EXO 1745-248 offers the opportunity of adding
an additional link to the evolutionary chain connecting LMXBs to MSPs.
In fact, from the analysis of the data currently available we can
speculate that EXO1745-248 is experiencing the very early phase of the
mass accretion stage, when an expanding star (a sub giant branch
object) is filling its Roche Lobe and transferring material that
eventually spins-up the neutron star.  Indeed, the few outbursts in
the X-ray occurring during this stage unambiguously indicate that
heavy mass accretion on the neutron star is taking place.  As time
passes, the mass accretion rate will decrease and the system will
enter into the phase characterized by a cyclic alternation between
accretion- and rotation-powered emission, as observed in the case of
MSP-M28I \citep{papitto13}.  Thus, in the next future EXO1745-248 is
expected to possibly start swinging between the LMXB and the MSP
states.  At later stages, when the neutron star has been sufficiently
re-accelerated, the emission is powered by the rotating magnetic
field, preventing further significant mass accretion. The system will
be observable as a redback, i.e., a radio MSP with a highly perturbed
companion, with the Roche Lobe filled and some material still falling
toward the neutron star.  Indeed, the prototype of this class of
objects is MSP-A in NGC 6397 (\citealp{fe01}; see also
\citealp{burderi02}), which shows a light curve dominated by tidal
distortions (in agreement with the fact that the companion is a
bloated, highly deformed star), prominent H$\alpha$ lines (confirming
the existence of diffuse gas outside the companion Roche Lobe;
\citealp{sabbi03}), and peculiar chemical patterns (as expected for a
deeply peeled star; \citealp{mucciarelli13}).  The final fate will be
that of a ``canonical MSP'', i.e., a NS in a binary system with a WD
companion (the stripped core of the donor star)\footnote{Note that
  unexpected features have been observed also in this case, the most
  notable example being the optical variability of the WD companion to
  the pulsar 6752A in NGC\,6752 \citep{cocozza06}.}.

\acknowledgments We warmly thank the referee, Craig Heinke, for 
the careful reading of the manuscript and useful suggestions that
help improving the paper.
This research is part of the project {\it Cosmic-Lab}
(web site: http://www.cosmic-lab.eu) funded by the European Research
Council (under contract ERC-2010-AdG-267675).

\end{document}